\documentclass[conference]{IEEEtran}\IEEEoverridecommandlockouts

\usepackage{graphicx, amssymb, amsmath, subfigure, setspace, bm, times}

\newtheorem{thm}{Theorem}
\newtheorem{prop}{Proposition}

\newtheorem{defi}{Definition}
\newtheorem{exam}{Example}

\def\squareforqed{\hbox{\rlap{$\sqcap$}$\sqcup$}}
\def\qed{\ifmmode\squareforqed\else{\unskip\nobreak\hfil
\penalty50\hskip1em\null\nobreak\hfil\squareforqed
\parfillskip=0pt\finalhyphendemerits=0\endgraf}\fi}

\newcommand{\ZZ}{{\mathbb Z}}
\newcommand{\RR}{{\mathbb R}}
\newcommand{\CC}{{\mathbb C}}

\newcommand{\FF}{{\mathbb F}}

\newcommand{\G}{\boldsymbol{G}}
\newcommand{\x}{\boldsymbol{x}}
\newcommand{\y}{\boldsymbol{y}}
\newcommand{\z}{\boldsymbol{z}}
\newcommand{\w}{\boldsymbol{w}}
\newcommand{\calE}{\mathcal{E}}
\newcommand{\calD}{\mathcal{D}}

\newcommand{\SNR}{{\sf SNR}}
\newcommand{\bI}{\boldsymbol{I}}

\begin{document}
%
\title{An Algebraic Approach to \\ Physical-Layer Network Coding}

\author{\IEEEauthorblockN{Chen Feng}
 \IEEEauthorblockA{Dept. of Electrical and Computer Eng.\\
 University of Toronto, Canada\\
 cfeng@eecg.utoronto.ca\vspace{-3ex}}
 \and
 \IEEEauthorblockN{Danilo Silva}
 \IEEEauthorblockA{School of Electrical and Computer Eng.\\
 State University of Campinas, Brazil\\
 danilo@decom.fee.unicamp.br\vspace{-3ex}}
 \and
 \IEEEauthorblockN{Frank R. Kschischang}
 \IEEEauthorblockA{Dept. of Electrical and Computer Eng.\\
 University of Toronto, Canada\\
 frank@comm.utoronto.ca\vspace{-3ex}}
 \thanks{The work of D. Silva was supported by FAPESP, Brazil.}
 }

\maketitle

\begin{abstract}
The problem of designing new physical-layer network coding (PNC) schemes via
lattice partitions is considered. Building on a recent work by
Nazer and Gastpar,
who demonstrated its asymptotic gain using information-theoretic tools, we take
an algebraic approach to show its potential in non-asymptotic settings.
We first relate Nazer-Gastpar's approach to the fundamental theorem of finitely generated modules over a principle ideal domain.
Based on this connection, we generalize their code construction and simplify their encoding and decoding methods. This not only provides a
transparent understanding of their approach, but more importantly, it opens up the opportunity to design efficient and practical PNC schemes.
Finally, we apply our framework for PNC to a Gaussian relay
network and demonstrate its advantage over conventional PNC schemes.
\end{abstract}

\IEEEpeerreviewmaketitle

\section{Introduction}

Physical-layer network coding (PNC) was proposed by Zhang {\em et al.}
\cite{ZLL06} to embrace interference in wireless networks. In a nutshell, each
relay in the network maps an interfering signal into an XOR combination of
simultaneously transmitted codewords. Surprisingly, this simple scheme doubles
the throughput of a two-way relay channel compared to traditional transmission
schemes \cite{ZLL06}.
Due to its remarkable potential, PNC has received considerable research
attention in recent years, with a particular focus on two-way relay systems \cite{KPT09}.

In a recent work \cite{NG09}, Nazer and Gastpar extended PNC from two-way relay
systems to general network scenarios. Their approach allows
each relay to map an interfering signal into some linear combination of transmitted messages
over a large prime field $\FF_p$. The underlying codes are based on lattice partitions whose algebraic structure makes this mapping reliable and efficient. They demonstrated its advantage over classical relaying strategies in various network scenarios. However,
their approach is essentially information-theoretic, as
they applied Loeliger's type A construction of random lattice ensembles \cite{Loe97} which
requires both the
field size $p$ and the codeword length to be sufficiently large.

Building on the theoretical insights of \cite{NG09}, we
extend their framework towards the
design of {\em efficient and practical} PNC schemes.
First, instead of using
Loeliger's type A construction to design asymptotically good lattice partitions, we
investigate a general design question: {\em Which class of lattice partitions is naturally suited for PNC?}

To answer this question, we apply
algebraic, rather than information-theoretic, tools. In particular, we show
that this question is closely related to the well-known fundamental theorem of
finitely generated modules over a principle ideal domain (PID). An important
consequence is that a large class of lattice partitions
has a vector space structure. This desirable property makes them well suited
for PNC, as a system of coset representatives can
then be used as codewords for PNC naturally and efficiently.


Second, we provide a sufficient condition for lattice partitions to have a vector space structure. This condition generalizes the code construction used in \cite{NG09}, leading to a large design space for efficient and practical PNC schemes. We then present encoding and decoding methods for our generalized code construction. One may expect that the generalized code construction requires more complicated encoder and decoder. Interestingly, we show that the encoder and decoder for our generalized code construction can be made even simpler than those proposed in \cite{NG09}. This is achieved by making use of the Smith normal form, another version of the fundamental theorem of finitely generated modules over a PID.

Our generalized code construction together with encoding and decoding methods provide an algebraic framework for PNC. To demonstrate its potential, we first revisit Nazer-Gastpar's approach using our algebraic framework. This leads to a more transparent understanding of their code construction, encoding and decoding. In particular, we show that a larger finite field can be obtained almost for free by setting the prime $p \equiv 3$ mod $4$ in their code construction.

As another application of our algebraic framework, we present a concrete design example for practical PNC schemes using the design space defined by our framework. We specify the code construction, encoding and decoding methods based on signal codes. Our simulation results suggest that PNC schemes using a $100$-dimensional signal code outperform conventional PNC schemes significantly. This confirms that PNC via lattice partitions indeed has a clear advantage even in non-asymptotic settings.

Proofs are generally omitted due to space constraints.

\section{Physical-Layer Network Coding}\label{sec:PNC}

\subsection{Problem Formulation}

As observed in \cite{NG09}, the cornerstone of a PNC scheme can be abstracted as the following problem of computing linear functions over Gaussian multiple-access channels (MAC).
\begin{figure}
\centering
\vspace{-3mm}
\includegraphics[width=0.45\textwidth]{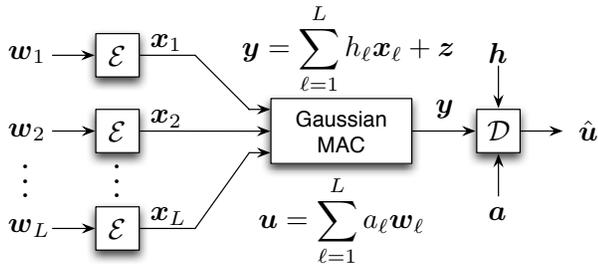}
\vspace{-4mm}
\caption{Computation over a Gaussian MAC.}
\label{fig:mac}
\vspace{-3mm}
\end{figure}

Each transmitter (indexed by $\ell = 1,\ldots, L$) is equipped with an identical encoder $\calE: \FF_q^k \to \CC^n$ that maps a message vector $\w_\ell \in \FF_q^k$ to a signal vector $\x_\ell = \mathcal{E}(\w_\ell) \in \CC^n$ satisfying the power constraint $\frac{1}{n} \| \x_\ell \|^2 \leq \SNR$. The \emph{rate} of the encoder, in bits per complex dimension, is defined as $(k \log q)/n$. For simplicity, we assume that each $\w_\ell$ is uniformly distributed in $\FF_q^k$ and independent of each other.

The receiver observes a noisy linear combination of the transmitted signal vectors through a Gaussian MAC channel:
\[
\y = \sum_{\ell = 1}^L h_{\ell} \x_\ell + \z,
\]
where $h_1,\ldots,h_L \in \CC$ are the channel coefficients, $\z \sim \mathcal{CN}(0, \bI_n)$ is a circularly-symmetric jointly-Gaussian complex random vector, and $\bI_n$ is the $n \times n$ identity matrix. Let $\boldsymbol{h} \triangleq (h_{1}, \ldots, h_{L})$ denote the channel coefficient vector. We assume that $\boldsymbol{h}$ is known at the receiver, but not at the transmitters.

The goal of the receiver is to reliably compute some linear combination of transmitted message vectors. Specifically, the receiver first selects a (finite-field) coefficient vector $\boldsymbol{a} \triangleq (a_{1}, \ldots, a_{L}) \in \FF_q^L$ based on $\boldsymbol{h}$. Then, it attempts to decode the linear combination $\boldsymbol{u} = \sum_{\ell = 1}^L a_{\ell} \w_\ell$ from the channel output $\y$ according to a decoder $\mathcal{D}_{\boldsymbol{h},\boldsymbol{a}}: \CC^n \to \FF_q^k$. Assume that $\boldsymbol{h}$ and $\boldsymbol{a}$ are fixed, and let $\hat{\boldsymbol{u}} = \mathcal{D}_{\boldsymbol{h}, \boldsymbol{a}}(\y)$. We say that a decoding error occurs if $\hat{\boldsymbol{u}} \ne \boldsymbol{u}$. The probability of error of the decoder is given by $\Pr(\hat{\boldsymbol{u}} \ne \boldsymbol{u})$.

Following \cite{NG09}, we say that a \emph{computation rate} $R$ is \emph{achievable} if, for any $\epsilon > 0$, any $\delta > 0$, and sufficiently large $n$, there exists an encoder $\calE$ with rate at least $R-\delta$ and a corresponding decoder $\calD_{\boldsymbol{h},\boldsymbol{a}}$ with probability of error less than $\epsilon$.

\subsection{Nazer-Gastpar's Approach}

Nazer and Gastpar proposed a PNC scheme \cite{NG09} based on {\em nested lattice codes}, which achieves a computation rate
\[
R(\boldsymbol{h}, \boldsymbol{a}) = \max \left\{ \log_2\left( \left( \| \boldsymbol{a} \|^2 - \frac{\mbox{SNR}|\boldsymbol{h} \boldsymbol{a}^{\dagger}|}{1 + \mbox{SNR} \| \boldsymbol{h} \|^2}  \right)^{-1} \right), 0 \right\},
\]
where the coefficient vector $\boldsymbol{a} \in \{\ZZ + i \ZZ \}^L$ and $\boldsymbol{a}^{\dagger}$ is the conjugate transpose of vector $\boldsymbol{a}$. Note that the vector $\boldsymbol{a}$ here is not a vector over a finite field. We will explain later how to interpret $\boldsymbol{a}$ as a finite-field vector.

Their approach is essentially information-theoretic, which relies on the existence of good nested lattice codes of infinitely high dimension. The objective of this paper is to design practical PNC schemes that can achieve a desired computation rate with low probability of error.

\section{An Algebraic Approach to PNC}\label{sec:framework}

Building on the theoretical insights of \cite{NG09}, we
extend their framework towards the
design of efficient and practical PNC schemes.
To this end, we apply
algebraic, rather than information-theoretic, tools.

\subsection{Preliminaries}

We briefly review some definitions and useful results related to our generalized code construction.
A more detailed treatment can be found in \cite{CS99}.

Let $\omega$ be a complex number such that the ring of integers $\ZZ[\omega] \triangleq \{ a + b \omega | a, b \in \ZZ \}$ is a principle ideal domain (PID). Two well known examples are Gaussian integers $\ZZ[i]$ and Eisenstein integers $\ZZ[(-1 + i \sqrt{3})/2]$. For ease of presentation, let $R$ denote $\ZZ[\omega]$ which is a PID. An {\em $R$-lattice} is defined as follows.
\begin{defi}[$R$-Lattices]
An $n$-dimensional {\em $R$-lattice} $\Lambda$ is defined by
a set of $n$ linearly independent (row) vectors
$\boldsymbol{g}_1, \ldots, \boldsymbol{g}_n$ in $\CC^{m}$ ($m \ge n$).
The lattice $\Lambda$ is composed of all $R$-linear combinations of the basis vectors:
	\[
	\Lambda = \{ r_1\boldsymbol{g}_1 + \cdots + r_n\boldsymbol{g}_n | r_1, \ldots, r_n  \in R \}.
	\]
	Equivalently,
	\[
	\Lambda = \{ \boldsymbol{r} \G_{\Lambda} | \boldsymbol{r} = (r_1, r_2, \ldots, r_n) \in
{R}^n \},
	\]
	where $\G_{\Lambda} \in \CC^{n \times m}$ is called a {\em generator
matrix} for $\Lambda$ with $\boldsymbol{g}_i$ as its $i$th row.
\end{defi}

Unless otherwise specified, we will assume that $m = n$ in this paper. An {\em $R$-sublattice} $\Lambda'$ of $\Lambda$ is a subset of $\Lambda$ which is itself an $R$-lattice.
The $R$-lattices and $R$-sublattices defined as above are
precisely $R$-modules and $R$-submodules. Hence, the set of
all cosets of $\Lambda'$ in $\Lambda$, denoted by $\Lambda / \Lambda'$
(also referred to as a {\em lattice partition}), forms a quotient $R$-module. If we take exactly one element from each coset, we obtain {\em a system of coset representatives}
for the partition $\Lambda / \Lambda'$.

The number of cosets of $\Lambda'$ in $\Lambda$ is called the {\em index} of $\Lambda'$ in $\Lambda$
and is denoted by $|\Lambda : \Lambda'|$. In this paper, we only consider the case when the index $|\Lambda : \Lambda'|$ is finite, which means the lattice $\Lambda$ is of the same dimension as the sublattice $\Lambda'$. Note that the lattice partition $\Lambda / \Lambda'$ of $|\Lambda : \Lambda'| < \infty$ is actually a {\em finitely generated torsion} $R$-module.

\subsection{Generalized Code Construction}

We are particularly interested in the condition under which a lattice partition
$\Lambda / \Lambda'$ forms a vector space over a finite field $\FF_q$. In this case, a system of coset representatives can be used naturally as codewords for PNC. To address this question,
we introduce the fundamental theorem
of finitely generated modules over a PID.
\begin{thm}[\cite{DF04}]\label{thm:fundamental}
	Let $R$ be a PID and let $M$ be a finitely generated torsion $R$-module. Then
	\[
	M \cong R/(r_1) \oplus R/(r_2) \oplus \cdots \oplus R/(r_k)
	\]
	for some integer $k > 0$ and nonzero elements $r_1, \ldots, r_k$ of $R$ which are not units in $R$ and which satisfy the divisibility relations $r_1 | r_2 | \cdots | r_k$. The ideal $(r_k)$ is the {\em annihilator} of $M$ defined by $\mbox{Ann}(M) = \{ r \in R | r m = 0, \ \mbox{for all} \ m \in M \}$.
Let $r$ be a nonzero, nonunit element of $R$. Suppose the factorization of $r$ into distinct prime powers in $R$ is
	\[
	r = u p_1^{\beta_1} p_2^{\beta_2} \cdots p_s^{\beta_s}
	\]
	where $u$ is a unit. Then $R/(r)$ can be further decomposed as
	\[
	R/(r) \cong R/(p_1^{\beta_1}) \oplus R/(p_2^{\beta_2}) \oplus \cdots \oplus R/(p_s^{\beta_s}).
	\]
\end{thm}

We next provide an algorithm to decompose the lattice partition $\Lambda / \Lambda'$. 	
Let $\G_{\Lambda}$ and $\G_{\Lambda'}$ be
the generator matrices for the lattice $\Lambda$ and sublattice $\Lambda'$. Suppose $\G_{\Lambda'} = \boldsymbol{J} \G_{\Lambda}$,
where $\boldsymbol{J}$ is an $n \times n$ matrix with entries from $R$. Using elementary row and column operations over $R$, the matrix $\boldsymbol{J}$ can be put into the diagonal form $\boldsymbol{D} = \mbox{diag}(1,\ldots, 1, d_1, \ldots, d_k)$ (called the {\em Smith normal form} for $\boldsymbol{J}$) for some integer $0 < k \le n$, and nonzero, nonunit elements $d_1, \ldots, d_k$ satisfying $d_1 | d_2 | \cdots | d_k$. The ideal $(d_k)$ is the annihilator of $\Lambda / \Lambda'$. As one may expect,
\[
\Lambda / \Lambda' \cong R/(d_1) \oplus R/(d_2) \oplus \cdots \oplus R/(d_k).
\]
In particular, if $d_k$ can be factored into distinct primes $p_i$ ($d_k = u p_1 p_2 \cdots p_s$) with the index $|R:(p_i)| = q$ for all $i = 1, \ldots, s$, then
\[
R/(d_k) \cong R/(p_1) \oplus R/(p_2) \oplus \cdots \oplus R/(p_s) \cong \FF_q^{s}.
\]
Since $d_i | d_k = u p_1 p_2 \cdots p_s$, it follows that $R/(d_i) \cong \FF_q^{s_i}$ for some $s_i$. Hence, we have the following theorem
	\begin{thm}\label{thm:vector}
		Let $\Lambda / \Lambda'$ be a lattice partition of $R$-lattices.
		If the annihilator $\mbox{Ann}(\Lambda / \Lambda') = (p_1 p_2 \cdots p_s)$, where the $p_i$ are distinct primes in $R$ satisfying $|R : (p_i)| = q$, then $\Lambda / \Lambda'$ is isomorphic to some vector space over the finite field $\FF_q$.
	\end{thm}
	\begin{exam}\label{ex:pi}
		Let $\Lambda / \Lambda'$ be a lattice partition of $R$-lattices with $\G_{\Lambda'} = \boldsymbol{J} \G_{\Lambda}$. Suppose $\mbox{Ann}(\Lambda / \Lambda') = (\pi)$ for some prime $\pi$ in $R$, that is, $\pi \Lambda \subseteq \Lambda'$. Then it follows immediately that the matrix $\boldsymbol{D}$ satisfies $d_1 = \cdots = d_k = \pi$ up to units. Hence, 	 there exist invertible matrices $\boldsymbol{P} \in R^{n \times n}$ and $\boldsymbol{Q} \in R^{n \times n}$ over $R$
			such that
			\begin{equation}\label{eq:D}
			\boldsymbol{P} \boldsymbol{J} \boldsymbol{Q} = \bar{\boldsymbol{D}} = \left[ \begin{array}{cc} \pi \boldsymbol{I}_k &  \boldsymbol{0}_{k \times (n - k)}  \\   \boldsymbol{0}_{(n - k) \times k} & \boldsymbol{I}_{n-k} \end{array} \right].
		\end{equation}
		Now we have
		\[
		\boldsymbol{P} \G_{\Lambda'} = \boldsymbol{P} \boldsymbol{J} \boldsymbol{Q} \boldsymbol{Q}^{-1} \G_{\Lambda} = \bar{\boldsymbol{D}} \boldsymbol{Q}^{-1} \G_{\Lambda}.
		\]
		Since the matrices $\boldsymbol{P}$ and $\boldsymbol{Q}$ are invertible over $R$, we can view $\boldsymbol{Q}^{-1} \G_{\Lambda}$ and $\boldsymbol{P} \G_{\Lambda'}$ as new generator matrices for the lattice $\Lambda$ and sublattice $\Lambda'$. In other words, we can assume, without loss of generality, that $\G_{\Lambda'} = \bar{\boldsymbol{D}} \G_{\Lambda}$ is this case.
	\end{exam}

	Theorem~\ref{thm:vector} provides a class of PNC-{\em compatible} lattice partitions, since a system of coset representatives can be used as codewords for PNC naturally and efficiently.
This code construction generalizes the nested lattice codes used in \cite{NG09}, leading to a larger design space for efficient and practical PNC schemes.  One such example will be given in Sec.~\ref{sec:signal} to demonstrate the usefulness of this generalized code construction.

\subsection{Encoding and Decoding}\label{sec:endecoder}

\newcommand{\Fq}{\FF_q}

We now propose explicit encoding and decoding methods for our generalized code construction. These methods are similar to those in \cite{NG09} but are simpler to implement due to our use of the Smith normal form as discussed above.



For ease of presentation, we focus on the special case given in Example~\ref{ex:pi} when $\G_{\Lambda'} = \bar{\boldsymbol{D}} \G_{\Lambda}$. The extension to more general cases is straightforward.

\renewcommand{\hom}{\varphi}
\newcommand{\mat}[1]{\begin{bmatrix} #1 \end{bmatrix}}
\newcommand{\enc}{\varphi^{-1}}

We start by exhibiting an explicit isomorphism between $\Fq^k$ and $\Lambda/\Lambda'$.
Let $\sigma: R \to R/\pi R \leftrightarrow \Fq$ be a surjective ring homomorphism,
and let us extend it to an $R$-module homomorphism\footnote{We can make $\Fq^k$ into an $R$-module by defining the action of $R$ on $\Fq^k$ as $a \cdot w = \sigma(a) w$.} $\sigma: R^k \to \Fq^k$ by applying it component-wise.
Let $\hom: \Lambda \to \Fq^k$ be defined by
  $\hom(\boldsymbol{\lambda}) \triangleq \sigma\left(\boldsymbol{\lambda} \boldsymbol{G}_\Lambda^{-1} \mat{\boldsymbol{I}_k \\ \boldsymbol{0}}\right)$.

\medskip
\begin{prop}
  The map $\hom: \Lambda \to \Fq^k$ is a surjective $R$-module homomorphism with $\ker \hom = \Lambda'$.
\end{prop}
\medskip

Let $\sigma^{-1}: \Fq^k \to R^k$ be some injective map such that $\sigma(\sigma^{-1}(\boldsymbol{w}))=\boldsymbol{w}$, for all $\boldsymbol{w} \in \Fq^k$. Similarly, let $\hom^{-1}: \Fq^k \to \Lambda$ be some injective map such that $\hom(\hom^{-1}(\boldsymbol{w}))=\boldsymbol{w}$, for all $\boldsymbol{w} \in \Fq^k$. Suitable choices for $\hom^{-1}$ are given by $\hom^{-1}(\boldsymbol{w}) = \sigma^{-1}(\boldsymbol{w})\mat{\boldsymbol{I}_k & \boldsymbol{B}}\boldsymbol{G}_\Lambda$, for any $\boldsymbol{B} \in \Fq^{k \times (n-k)}$. In the following, we will use
\begin{equation}\label{eq:map1}
  \hom^{-1}(\boldsymbol{w}) = \sigma^{-1}(\boldsymbol{w})\mat{\boldsymbol{I}_k & \boldsymbol{0}}\boldsymbol{G}_\Lambda
\end{equation}
unless otherwise mentioned.

%
%


\medskip
\subsubsection{Construction of the encoder}
The encoder $\calE$ consists of the map $\enc$, together with a dither $\boldsymbol{v} \in \CC^n$ and a shaping operation.


Given a message $\boldsymbol{w} \in \Fq^k$, we first compute a lattice point $\enc(\boldsymbol{w}) \in \Lambda$. We then add a dither $\boldsymbol{v} \in \CC^n$ to obtain a vector $\enc(\boldsymbol{w}) + \boldsymbol{v}$. The purpose of the dither $\boldsymbol{v}$ is explained in \cite{NG09}. Since the norm of the vector $\enc(\boldsymbol{w}) + \boldsymbol{v}$ may be large, we further add a lattice point $\boldsymbol{\lambda}'$ of the sublattice $\Lambda'$ in order to reduce the power consumption. This operation is called ``shaping''. As an example, one may choose
\[
\boldsymbol{\lambda}' = - Q_{\Lambda'}(\enc(\boldsymbol{w}) + \boldsymbol{v}),
\]
where $Q_{\Lambda'}(\cdot): \CC^n \to \Lambda'$ is a {\em lattice quantizer} that sends a point in $\CC^n$ to the nearest lattice point of $\Lambda'$.

Thus, the encoder $\calE$ is given by
\begin{equation}\label{eq:encoder}
\x = \calE(\boldsymbol{w}) = \enc(\boldsymbol{w}) + \boldsymbol{v} + \boldsymbol{\lambda}',
\end{equation}
where $\boldsymbol{\lambda}' \in \Lambda'$. Essentially, the encoder $\calE$ provides
an one-to-one mapping between the vector space $\FF_q^k$ and a system of coset representatives for $\Lambda / \Lambda'$ shifted by $\boldsymbol{v}$.

\subsubsection{Construction of the decoder}

The decoder $\mathcal{D}_{\boldsymbol{h},\boldsymbol{a}}$ consists of an affine operator $g(\cdot)$
and a lattice quantizer $Q_{\Lambda}$ for the lattice $\Lambda$, as well as the map $\hom$.

The affine operator $g(\cdot)$ is defined as follows
\begin{equation}\label{eq:mse}
g(\boldsymbol{y}) = \alpha \boldsymbol{y} - \sum_{\ell = 1}^L a_\ell\boldsymbol{v}_\ell,
\end{equation}
where $\boldsymbol{v}_\ell \in \CC^n$ is the dither for user $\ell$, and $\alpha \in \CC$ is a scaling factor specified in \cite{NG09} to minimize the probability of error.
Applying (\ref{eq:encoder}) in (\ref{eq:mse}), we have
\[
	g(\boldsymbol{y}) =  \sum_{\ell = 1}^L a_\ell\left(\enc(\w_\ell) + \boldsymbol{\lambda}_\ell'\right) + \boldsymbol{n},
\]
where $\boldsymbol{n} \triangleq \sum_{\ell = 1}^L (\alpha h_\ell - a_\ell) \x_\ell + \alpha \z$. Note that $g(\boldsymbol{y}) - \boldsymbol{n} \in \Lambda$.

Let $\boldsymbol{u} = \sum_{\ell = 1}^L a_\ell\w_\ell$ be a linear combination of message vectors. We have that $\hom(g(\boldsymbol{y})-\boldsymbol{n}) = \sum_{\ell = 1}^L a_\ell\w_\ell = \boldsymbol{u}$. Thus, the receiver may attempt to decode by computing
\begin{equation}\nonumber
  \hat{\boldsymbol{u}} = \mathcal{D}_{\boldsymbol{h},\boldsymbol{a}}(\boldsymbol{y}) = \hom(Q_\Lambda(g(\boldsymbol{y}))) = \boldsymbol{u} + \hom(Q_\Lambda(\boldsymbol{n}))
\end{equation}
Note that an error occurs only if $Q_\Lambda(\boldsymbol{n}) \neq \boldsymbol{0}$.


More concretely, the decoder first finds
\begin{equation}\label{eq:decoder}
\hat{\boldsymbol{r}} = \arg_{\boldsymbol{r} \in R^n}\min \| g(\boldsymbol{y}) - \boldsymbol{r} \G_{\Lambda} \|^2
\end{equation}
and then computes
$\hat{\boldsymbol{u}} = \sigma(\hat{r}_1, \ldots, \hat{r}_k)$,
where $(\hat{r}_1, \ldots, \hat{r}_k)$ denotes the first $k$ entries in $\hat{\boldsymbol{r}}$.

\section{Applications of the Algebraic Approach} \label{sec:apply}

Our generalized code construction together with encoding and decoding methods provide an algebraic framework for PNC. To demonstrate its potential, we provide two applications in this section.

\subsection{Nazer-Gastpar Revisited}

First, we revisit Nazer-Gastpar's approach using our algebraic framework, leading to a more transparent understanding.

\subsubsection{Code construction}
The nested lattice codes used in \cite{NG09} are constructed as follows. First, pick an $n$-dimensional real lattice $\Lambda'_r$ with a generator matrix $\G_{\Lambda'_r} \in \RR^{n \times n}$ that is simultaneously good for covering, quantization, and AWGN channel coding. The corresponding $\ZZ[i]$-lattice $\Lambda'$ is given by
\[
\Lambda' = \{ \boldsymbol{r} \G_{\Lambda'_r} = \mbox{Re}\{\boldsymbol{r}\} \G_{\Lambda'_r} + i \ \mbox{Im}\{\boldsymbol{r}\} \G_{\Lambda'_r} | \boldsymbol{r} \in \ZZ[i]^n \}.
\]
Second, draw a matrix $\boldsymbol{B} \in \ZZ^{k \times (n-k)}$
with every element $b_{ij}$ chosen i.i.d. according to the uniform distribution over $\{ 0, 1, \ldots, p-1\}$ where $p$ is a prime in $\ZZ$. Construct another $\ZZ[i]$-lattice $\Lambda$ as follows
\[
\Lambda = \{ \boldsymbol{r} \G_{\Lambda_r} = \mbox{Re}\{\boldsymbol{r}\} \G_{\Lambda_r} + i \ \mbox{Im}\{\boldsymbol{r}\} \G_{\Lambda_r} | \boldsymbol{r} \in \ZZ[i]^n \},
\]
where
\[
\G_{\Lambda_r} \triangleq {p}^{-1} \left[ \begin{array}{cc}  \boldsymbol{I}_k &  \boldsymbol{B}_{k \times (n - k)}  \\   \boldsymbol{0}_{(n - k) \times k} & p \boldsymbol{I}_{n-k} \end{array} \right] \G_{\Lambda'_r}.
\]
Let $\G_{\Lambda'_r} = \boldsymbol{J} \G_{\Lambda_r}$. We have
\[
\boldsymbol{J} =  \left[ \begin{array}{cc} p \boldsymbol{I}_k &  -\boldsymbol{B}_{k \times (n - k)}  \\   \boldsymbol{0}_{(n - k) \times k} & \boldsymbol{I}_{n-k} \end{array} \right].
\]
Hence, the $\ZZ[i]$-lattice $\Lambda'$ is indeed a sublattice of $\Lambda$, since entries of $\boldsymbol{J}$ are from $\ZZ[i]$. By Theorem~\ref{thm:fundamental}, we have $\Lambda / \Lambda' \cong (\ZZ[i]/(p))^k$. This result implies that if $p$ is also a prime in $\ZZ[i]$, then the field size is actually $p^2$ rather than $p$ as claimed in \cite{NG09}. In other words, we can get a larger finite field almost for free by setting the prime $p \equiv 3$ mod $4$, since such prime numbers are also primes in $\ZZ[i]$.

\subsubsection{Encoding}
The encoder $\calE$ in \cite{NG09} is given by
\[
\x = \calE(\boldsymbol{w}) = \enc(\boldsymbol{w}) + \boldsymbol{v} - Q_{\Lambda'}(\enc(\boldsymbol{w}) + \boldsymbol{v}),
\]
where the mapping $\enc$ is
\[\enc(\boldsymbol{w}) = {p}^{-1} \sigma^{-1}(\boldsymbol{w}) [  \boldsymbol{I}_k\  \boldsymbol{B}_{k \times (n - k)}] \G_{\Lambda'_r}.
\]
Note that our mapping is given in (\ref{eq:map1}). Although they are equivalent, our mapping requires few operations since it avoids the multiplication by $\boldsymbol{B}_{k \times (n - k)}$.

\subsubsection{Decoding}
The decoder $\calD_{\boldsymbol{h},\boldsymbol{a}}$ in \cite{NG09} consists of an affine operator $g(\cdot)$, a lattice quantizer $Q_{\Lambda}$ for the lattice $\Lambda$, another lattice quantizer $Q_{\Lambda'}$ for the sublattice $\Lambda'$, and a linear mapping $\phi^{-1}$ (similar to our $\hom$ but much more complex) given in \cite[Eq. (58)]{NG09}. Our decoder $\calD_{\boldsymbol{h},\boldsymbol{a}}$ described in (\ref{eq:decoder}) uses exactly the same affine operator $g(\cdot)$ and lattice quantizer $Q_{\Lambda}$, but avoids all the remaining calculations, which is quite beneficial in practice. This is achieved by making use of the Smith normal form of the matrix $\boldsymbol{J}$.

\subsection{An Example of Practical PNC Schemes}\label{sec:signal}

As another application of our algebraic framework, we present a concrete design example of PNC schemes. Recall that the design space defined by our algebraic framework  includes an $R$-lattice $\Lambda$ with sublattice $\Lambda'$, dithers $\boldsymbol{v}_\ell$ for each user $\ell$, a shaping operation, as well as a lattice decoder $\calD_{\Lambda}$.

\subsubsection{Code construction}
Nazer and Gastpar make use of existence of good (infinitely high dimensional) real lattices to produce a $\ZZ[i]$-lattice $\Lambda$ with sublattice $\Lambda'$, which is however very difficult to implement in practice. In contrast, we use existing practical, high coding gain lattices, such as signal codes \cite{SSF08}. Signal codes are a special class of $\ZZ[i]$-lattices whose generator matrix is given by
\[
\G_{\Lambda}^{k \times (k + m)} = \left[
\begin{array}{ccccccccc}
	1 & f_1  & \cdots & f_m & 0 &   \cdots & 0 & 0 \\
	0 & 1  & \cdots & f_{m - 1} & f_m &   \cdots & 0 & 0 \\
	 \vdots & \vdots & \cdots  & \vdots & \vdots & \cdots & \vdots & \vdots \\
		0 & 0 & \cdots & 0 & 0 & \cdots & f_{m-1} & f_m
	\end{array}  \right]
\]
where $f_i \in \CC$, for $i = 1, \ldots, m$. The sublattice $\Lambda'$ in our construction is chosen to be
$\G_{\Lambda'} = p \G_{\Lambda}$, where $p \equiv 3$ mod $4$ is a prime in both $\ZZ$ and $\ZZ[i]$.
Hence, we have $\Lambda / \Lambda' \cong \FF_{q}^k$, where $q = p^2$.

\subsubsection{Encoding}
To construct the encoder $\calE$,
we shall specify the dithers and the shaping operation. In fact, we set $\boldsymbol{v}_\ell = \boldsymbol{0}$ for $\ell = 1, \ldots, L$. In other words, we remove all the dithers. This is because the dither $\boldsymbol{v}_\ell$ is just a tactic to simplify some proof, but is not actually needed in practice \cite{For03}.

We apply the Tomlinson-Harashima shaping as suggested in \cite{SSF08}, which is a special case of the shaping operations defined in Sec.~\ref{sec:endecoder}. As a result, our encoder $\calE$ is identical to that used in \cite{SSF08}.


\subsubsection{Decoding}
The decoder $\mathcal{D}_{\boldsymbol{h}, \boldsymbol{a}}(\y)$ consists of an affine operator $g(\cdot)$, a Tomlinson-Harashima shaping operator, and a signal code decoder. The scaling factor $\alpha$ for $g(\cdot)$ is specified in \cite[Eq. (40)]{NG09}. The shaping operator is identical to that in our encoder $\calE$. By applying this operator, the heap-based stack decoder proposed in \cite{SSF08} can be used without any modification. We emphasize here this additional shaping operator does not involve any loss of information, as explained in Sec.~\ref{sec:endecoder}.

\section{Simulation Results}\label{sec:simulation}

To demonstrate the potential for PNC schemes proposed in Sec.~\ref{sec:signal}, we present simulation results for an illustrative network scenario. We then discuss several more elaborate designs that may achieve better performance in practice. The work along this line is in progress.

\begin{figure}[htb]
\centering
\vspace{-3mm}
\includegraphics[width=0.45\textwidth]{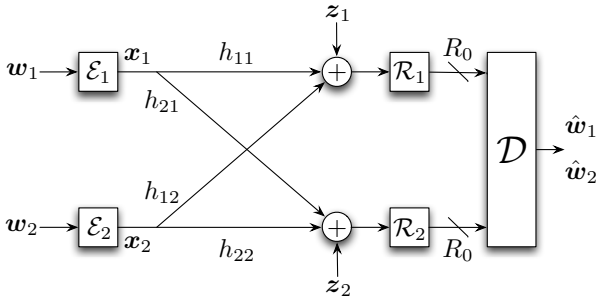}
\vspace{-4mm}
\caption{A canonical Gaussian relay network.}
\label{fig:network}
\vspace{-3mm}
\end{figure}

Similar to \cite{NG09}, we consider a canonical Gaussian relay network with two transmitters and a single decoder, as depicted in Fig.~\ref{fig:network}. Two relays in the network are connected to the decoder through rate-limited bit pipes. For the purpose of illustration, we
set $\G_{\Lambda'} = 3 \G_{\Lambda}$, resulting in a finite field $\FF_9$. The generator matrix for $\G_{\Lambda}$ is given in Sec.~\ref{sec:signal}, with parameters setting to $f_1 = 1.96 e^{i\pi/8}$, $f_2 = 0.98^2 e^{i \pi/4}$, and $m = 2$.

\begin{figure}[htb]
\centering
\vspace{-3mm}
\includegraphics[width=0.42\textwidth]{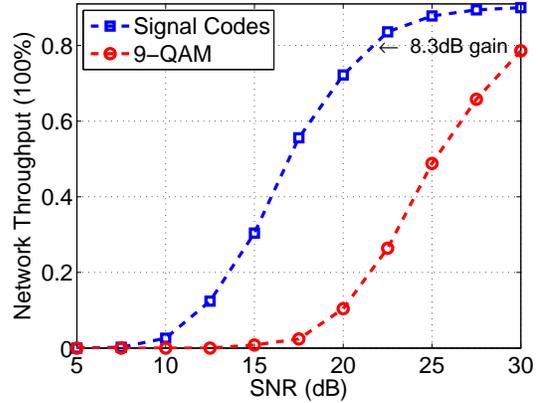}
\caption{Comparison of achievable throughput of PNC schemes using signal codes and
uncoded QAM.}
\label{fig:network1}
\vspace{-3mm}
\end{figure}

We evaluate the performance in terms of the network throughput. We focus on block Rayleigh fading channels in which the channel coefficients are independent $\mathcal{CN}(0,1)$ random variables, independent of the Gaussian noise. In our simulations, the data packet is set to have $100$ symbols ({\em i.e.,} $\boldsymbol{w}_\ell \in \FF_9^{100}$ for $\ell = 1, 2$). Fig.~\ref{fig:network1} shows the network throughput achieved by our PNC scheme under different {\sf SNR}s. For comparison, we also plot the curve of the baseline performance for PNC using uncoded $9$-QAM. Note that $9$-QAM is based on the lattice partition $\ZZ[i]^n / 3\ZZ[i]^n$ so it is again a very special case of our generalized code construction. 

From Fig.~\ref{fig:network1}, it is confirmed that PNC via lattice partitions indeed has a clear advantage over conventional PNC schemes even in non-asymptotic settings. With a simple $81 (= 9 \times 9)$ state signal code, it outperforms that using $9$-QAM by $8.3$~dB at $90\%$ throughput achievement.

To further improve the throughput, more elaborate designs are needed. For example, nested lattice shaping \cite{SSF08} may be used such that additional $1.53$~dB of shaping gain can be potentially obtained. Another candidate of practical, high coding gain lattices is low density lattice codes (LDLC) \cite{SFS06}. It has recently been reported that a $100$-dimensional LDLC using nested lattice shaping can be made only $3.6$~dB from the sphere bound \cite{KDL09}. This makes it very attractive to be used in our framework.

\section{Conclusion}\label{sec:conclude}

In this paper, we have followed the framework of Nazer
and Gastpar~\cite{NG09} towards the design of
PNC schemes via
lattice partitions for general network scenarios.
We have taken an algebraic approach, and have found a class
of PNC-compatible lattice partitions,
which generalizes the code construction in \cite{NG09}. We then developed encoding and
decoding methods for this generalized code construction. This not only provides a transparent
understanding of Nazer-Gastpar's approach, but more importantly, it opens up the opportunity
for practical PNC designs.
Finally, we have presented an illustrative
design example to demonstrate the potential of our algebraic approach.
We believe that we have merely scratched the surface of a potentially
rich and practically useful research area.

\bibliographystyle{IEEEtran}
\bibliography{main}

\end{document}